\documentclass[letterpaper, 10 pt, conference]{ieeeconf}  

\IEEEoverridecommandlockouts                              

\usepackage{graphicx} 
\usepackage{amsmath} 
\usepackage{amssymb}  
\usepackage{todonotes}
\usepackage{url}

\mathchardef\mhyphen="2D
\newcommand\fscore{\mathit{F\mhyphen score}}

\title{\LARGE \bf

One-class autoencoder approach for optimal electrode set-up identification in wearable EEG event monitoring*

}

\author{Laura M. Ferrari$^{1}$, Guy Abi Hanna$^{2\dagger}$, Paolo Volpe$^{2\dagger}$, Esma Ismailova$^{3}$, \\François Bremond$^{1}$ and Maria A. Zuluaga$^{2}$
\thanks{*This work has been partially supported by the Ville de Nice and the French government, through the UCAJEDI Investments in the Future project managed by the National Research Agency (ANR) (ANR-15-IDEX-01), and by the National Research Agency ANR JCJC OrgTex project (ANR-17-CE19-0006-01) }
\thanks{$^{1}$L.M. Ferrari and F. Bremond are with Stars team, INRIA 06902 Sophia Antipolis Cedex, France
        {\tt\small [laura.ferrari,
        francois.bremond]@inria.fr}}%
\thanks{$^{2}$G. Abi Hanna, P. Volpe and M.A. Zuluaga are with Data Science Department, EURECOM, 06410 Sophia Antipolis, France
        {\tt\small [abi-hanna,volpe,zuluaga]@eurecom.fr}}%
\thanks{$^{3}$E. Ismailova is with Department of Bioelectronics, Mines Saint-Étienne, 13541 Gardanne, France
        {\tt\small ismailova@emse.fr}}
\thanks{$\dagger$ Equal contribution}
}

\begin{document}

\maketitle
\thispagestyle{empty}
\pagestyle{empty}

\begin{abstract}
A limiting factor towards the wide real-life use of wearables devices for continuous healthcare monitoring is their cumbersome and obtrusive nature. This is particularly true for electroencephalography (EEG) recordings, which require the placement of multiple electrodes in contact with the scalp. In this work we propose to identify the optimal wearable EEG electrode set-up, in terms of minimal number of electrodes, comfortable location and performance, for EEG-based event detection and monitoring. By relying on the demonstrated power of autoencoder (AE) networks to learn latent representations from high-dimensional data, our proposed strategy trains an AE architecture in a one-class classification setup with different electrode set-ups as input data. The model performance is  assessed using the F-score. Alpha waves detection is the use case, through which we demonstrate that the proposed method allows to detect an alpha state from an optimal set-up. The so-called wearable configuration, consisting of electrodes in the forehead and behind the ear, is the chosen optimal set-up, with an average F-score of 0.78. Our results suggest that a learning-based approach can be used to enable the design and implementation of optimized wearable devices for real-life event related healthcare monitoring.  
\end{abstract}


\section{INTRODUCTION}
\begin{figure}[t]
    \centering
    \centerline{\includegraphics[width=0.9\columnwidth]{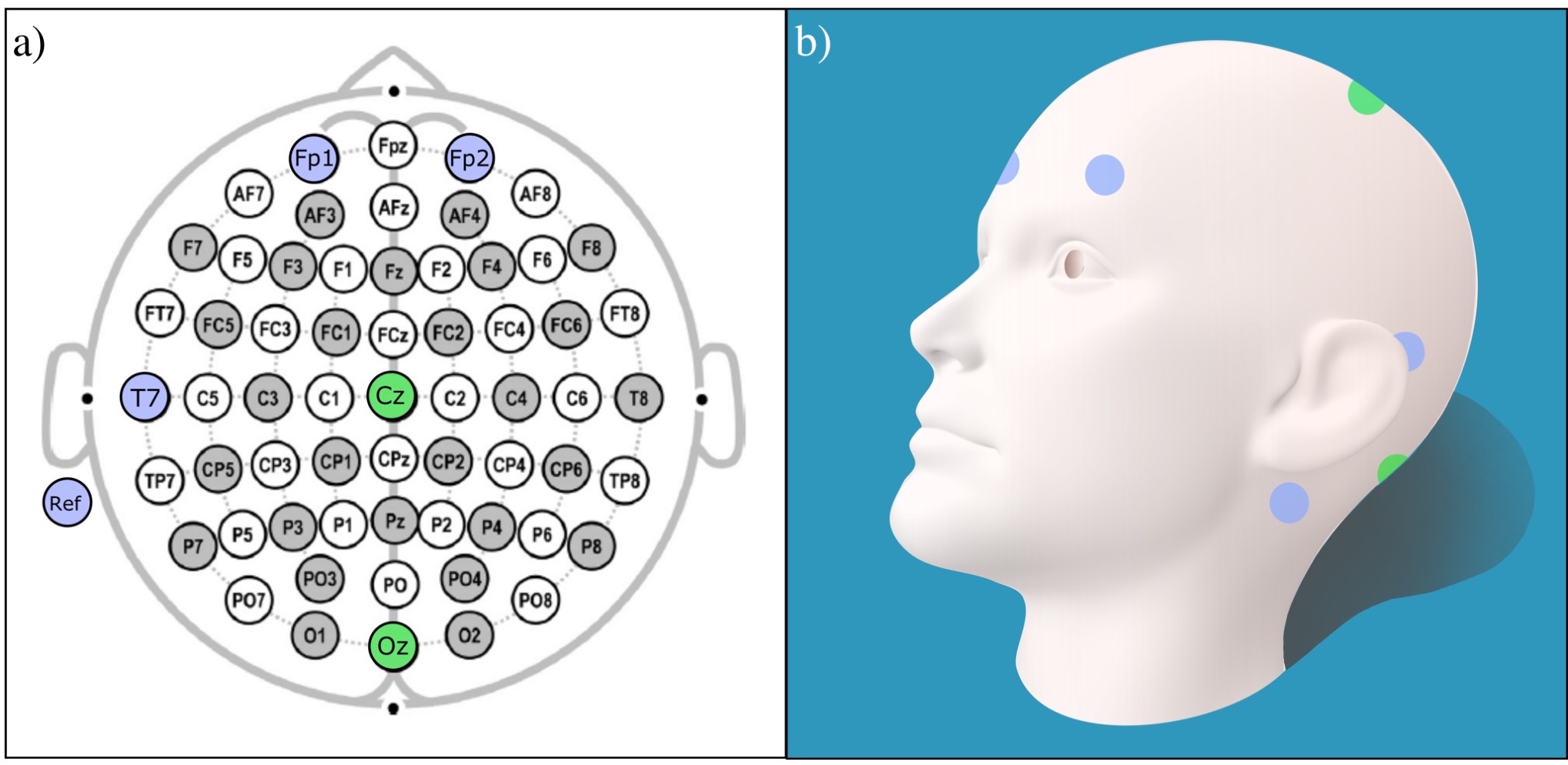}}
    \caption{The EEG headset. a) International 10-20 system with 32 electrodes (Gray circles) ~\cite{yoon2013} and the electrodes adopted in this study highlighted (light-blue, violet). b) 3D representation of the headset, with green electrodes (Cz-Oz) and the wearables channels, in violet, here investigated (the reference electrode is placed on the left mastoid bone for simplicity in this representation).}\label{fig:eeg} 
\end{figure}

 Electroencephalography (EEG) recording is the defacto approach to brain functions assessment with diagnostic or monitoring purposes (e.g. epilepsy, sleep studies). It is performed through electrodes placed along the scalp that non-invasively transduce the brain's electrical activity. The standard international system for electrodes placement with a configuration of 32 electrodes is depicted in Fig.~\ref{fig:eeg}(a). The need of such a dense electrodes' locations is a limiting factor in-view of real-life monitoring applications.
 
 Wearable EEG represents a promising solution to achieve ubiquitous monitoring~\cite{casson2008},
 for which there exist some commercial solutions 
 (e.g. Emotive cask\footnote{\url{https://www.emotiv.com/}}) relying on a simplified electrodes scheme. However, they have a cumbersome interface, in terms of materials and set-up of the electrodes. To overcome the limitations of traditional bulky and rigid materials, promising alternatives have been proposed in the field of epidermal~\cite{kim2011} and tattoo electronics~\cite{ferrari2020}. 
 Despite their seamless interface, the set-up still requires multiple electrodes in uncomfortable or non-discrete locations. An optimized electrodes set-up, i.e. minimum number in comfortably and discrete locations, for specific applications is still missing to achieve a realistic use of wearable EEG. 
 
In this work, we propose to use a deep neural network to identify the optimal electrode set-up to monitor a given state or condition from EEG recordings. 
 To this end, we model EEG recordings acquired through tattoo electrodes~\cite{ferrari2020b} as multi-variate time series. Under the hypothesis that collected electrophysiological signals are a representation of a latent condition, we train an autoencoder (AE) network to learn a model of the variability of such condition. To avoid the class imbalance problem during training~\cite{chambon2018}, we formulate our problem as a one-class classification one~\cite{pimentel2014}.
At inference time, the trained AE detects the presence or absence of the condition/state of interest in unseen data points. 
Using this configuration, we propose to alter the number of variables of the multi-variate time series, i.e. the EEG channels, to identify the optimal set-up that identifies the condition of interest in unseen data with acceptable performance. We investigate alpha waves detection, the most studied brain rhythm, as a use case to validate the proposed approach.

The remaining of this paper is organized as follows. Section~\ref{sec:alpha} introduces our use case, the alpha waves detection. Section~\ref{sec:method} describes the proposed method. In Section~\ref{sec:experiments} we present the experimental setup and the obtained results. The paper concludes with a summary of our contributions, a discussion of the related works and future perspectives.

\section{USE CASE: ALPHA WAVE DETECTION}\label{sec:alpha}
Alpha waves are a spontaneous brain activity that appears in the 8–12 Hz frequency band. They are induced by relaxation with closed eyes and abolished by eye opening or alerting (e.g. thinking, calculating)~\cite{teplan2002}. 
In relaxation or drowsiness, alpha activity is known to rise and, if sleep appears, the power of the lower frequency bands increases~\cite{teplan2002}. 
Alphas detection finds applications as indicators of sleepiness in high risk activities, as long-distance driving~\cite{jiao2020}, and as a marker of sleep depth~\cite{asyali2007}. 

Alphas typically arise in the occipital region and they are well visible in the EEG's CzOz channel, where a channel represents the acquisition from two electrodes. The CzOz channel is impracticable for compact and comfortable wearable devices, as it involves the whole back part of the head (Fig.~\ref{fig:eeg}). Considering that the cerebral activity arises inside the brain and it is spread all over the scalp surface, we hypothesize that it is possible to infer the presence or absence of alpha waves from other channels, through a learnt latent representation. In particular, we are interested in identifying a subset of locations of the 10-20 international system that are more realistic for wearable implementation, such as behind-the-ear (T7 location,  Fig.~\ref{fig:eeg}), which has been explored for seizure detection~\cite{gu2018}, or forehead EEG (Fp1-Fp2 electrodes, Fig.~\ref{fig:eeg}).

\section{METHOD}\label{sec:method}
This section first formulates the AE-based one-class classification problem (\ref{sec:oneclass}) for alpha wave detection. Next, we introduce the optimal electrode set-up selection strategy (\ref{sec:strategy}). The section concludes with a description of the implementation of the AE network (\ref{sec:network}). 
\subsection{AE-based One-class Classification of Alpha Waves}\label{sec:oneclass}
Let us denote
$\smash{\mathcal{T}=\left\lbrace \mathbf{x}_t\right \rbrace_{t\in T}}, \mathbf{x} \in \mathbb{R}^m$ is a multivariate time series, representing an EEG recording corresponding to $m$ channels, with each point $\mathbf{x}_t$ being an observation at a specific time $t$. 
%
One-class time-series classification trains a model under the assumption that the training data $\mathcal{T}$ comes from a single class, denoted the positive class. At inference time, the goal is to identify if unseen observations $\mathbf{\hat{x}}_t \notin \mathcal{T}$ belong to the positive class or not 
, under the assumption that $\mathbf{\hat{x}}_t$ belongs to the positive class if it is similar to the observations from $\mathcal{T}$, according to some (dis-)similarity metric. In this work, we consider alpha waves as the positive class since they represent the condition of interest.  

An AE is a neural network combining an encoder $E$ and a decoder $D$. The encoder part takes an input $X$ and maps it into a set of latent variables $Z$. The decoder maps from the latent space back into the input space as a reconstruction. The difference between the original input vector and its output is denoted the reconstruction error 
\begin{equation}
    \|X -AE(X)\|_2
\end{equation}
where $\|\cdot\|$ the L$_2$ norm, and 
\begin{equation}\label{eq:AE}
AE(X)=D(Z), \quad Z=E(X). 
\end{equation}
Trained with data $\mathcal{T}$ from the positive class, the AE estimates a model that captures the dynamics of such class~\cite{blazquez2020}. At inference time, the AE reconstructs well data similar to $\mathcal{T}$, while failing to do so with data that it has not encountered, thus resulting in large reconstruction errors. This error is used as a score to classify new points into the positive (low error) or negative class (high error).

To model the dependence between a current time point and previous ones it is common to define, at every $t$, a time window of length $K < |T|$ , i.e. $W_t=\left\lbrace \mathbf{x}_{t-K+1}, \ldots, \mathbf{x}_{t-1}, \mathbf{x}_{t}\right\rbrace$. This means that the original time-series $\mathcal{T}$ is transformed into a sequence of windows $\mathcal{W}=\smash{\left\lbrace W_t\right \rbrace_{t\in T}}$ to be used as training input. Raw electrophysiological signals, however, are complex and generally noisy~\cite{hakonen2015}. To avoid spurious effects linked to the nature of the data, we do not build the standard time windows $W$ of raw time points $\mathbf{x}_t$. Instead, we transform raw time point windows $W^{m\times K}\rightarrow W^{m\times L}$, by extracting a set of $L$ time- and frequency-domain features per time window (Table~\ref{table:features}).

\begin{table}[t]
\caption{Set of extracted features, $L=8$}
\label{table:features}
\centering
\begin{tabular}{|c||c|}
\hline
Type & Features\\
\hline
Time-domain & \multicolumn{1}{l|}{Mean, Standard deviation, Median, Minimum,} \\
& \multicolumn{1}{l|}{Maximum, Root-mean-square (RMS) }\\
Frequency-domain & \multicolumn{1}{l|}{Maximum Power Spectral Density (PSD), Mean} \\
& \multicolumn{1}{l|}{PSD}\\
\hline
\end{tabular}
\end{table}

\subsection{Optimal Electrode Set-up Selection}\label{sec:strategy}
AEs are good at extracting low-dimensional subspaces (latent spaces) representing the dynamics inside a high-dimensional dataset. The proposed method uses this property to identify the best set of EEG channels to use, i.e. the minimum set of comfortable and discrete channels, which is able to detect the presence/absence of an alpha state. We vary $m$, the number of input EEG channels, to train candidate models and assess their performance using an evaluation metric. We choose to use the F-score as it a well-suited evaluation metric for class imbalanced data, but any other performance measure could be used. 
\begin{equation}\label{eq:fscore}
    \fscore = \dfrac{2TP}{2TP + FN+FP}
\end{equation}
with TP denoting true positives, FP a time point misclassified as the positive class, and FN a false negative. 

In alpha waves detection, we expect a maximal performance using the CzOz channel as input channel, since it is the one normally used to measure alpha activity. We adopt it here as the reference. The optimal wearable design is chosen among all candidate models based on performance (closest to the reference), number of channels (the least the better) and comfort (at the least, avoid Cz).


\subsection{Network Implementation}\label{sec:network} We use an unsupervised AE-based topology, as the backbone architecture, which has proved superior performance in multivariate time series analysis~\cite{audibert2020}. 
The network is composed of a common encoder $E$ connected to two decoder networks $D_1$ and $D_2$: $AE_1(X) = D_1(Z), \,\, AE_2(X) = D_2(Z)$, with $Z$ as in Eq~\ref{eq:AE}. 
The network is trained using a two-phase adversarial training scheme to allow the AEs to learn how to amplify the reconstruction error as detailed in~\cite{audibert2020}. 
At inference time, the score of unseen data $\widehat{X}$ is estimated as a linear combination of the reconstruction error of the two AEs: 
\begin{equation}
  S(\widehat{X}) = \alpha \|\widehat{X}-AE_1(\widehat{X})\|_2 + \beta\|\widehat{X}-AE_2(AE_1(\widehat{X}))\|_2, 
\end{equation}
where $\alpha+\beta=1$ are two hyper-parameters that control sensitivity and specificity.

\section{EXPERIMENTS AND RESULTS}\label{sec:experiments}
This section describes the data (\ref{subsec:data}), the experimental setup  (\ref{subsec:setup}) and the results of the proposed method along with a comparison with other approaches (\ref{subsec:results}). 

\subsection{Data}\label{subsec:data}
EEG dataset was acquired as described in~\cite{ferrari2020b}. 
The tattoo electrodes were placed in T7, Cz, Oz, Fp1, and Fp2 locations (Fig.~\ref{fig:eeg}). A tattoo reference electrode (ref) was placed on the right mastoid bone, while the ground was located at the highest point of the head, near the Cz position.
For the alpha session, the participant, comfortably seated in an isolated room, was asked to close the eyes to produce alpha waves and, when requested, to open the eyes to stop their appearance. The non-alpha sessions had the same set-up with open eyes at all time. A total of 13 recordings were acquired, from which 9 have been used in this study, with a length of $\sim$2 minutes, accounting for 3.07M raw sample points. For each recording, the time points are labeled by an expert rater.

\subsection{Setup}\label{subsec:setup}
We assessed five different EEG channel configurations to identify the optimal electrode set-up. These are: 1) \textit{all}: T7Cz, Fp1Fp2, refCz, refT7, refOz, refFp1; 2) \textit{noCz}: Fp1Fp2, refT7, refOz; 3) \textit{wearable}: Fp1Fp2, refT7; 4) \textit{refT7}: 
and 5) \textit{Fp1Fp2}: also known as forehead EEG. We used the signals without any pre-processing. The dataset can be noisy and with typical EEG artefacts. We split the training and testing set over 6 different folds. The training set of the AE-based network only used positive class samples.
The optimal set-up was selected by estimating the average F-score (Eq.~\ref{eq:fscore}) on the test set, over the 6 folds. 
Our backbone network used a publicly available implementation\footnote{\url{https://github.com/robustml-eurecom/usad}} with $\alpha$=$\beta$=$0.5$, window size $K$=$4$ and the latent space dimension $|Z|$=$0.5m\cdot L$. 
 
We compared the performance of the proposed architecture with two classical machine learning approaches: a random forest (RF), with 500 trees and maximum depth of 5, and a Gradient Boosted Tree (GBT). Both coded in Python using the Scikit-learn implementation. We performed a 6-fold cross-validation by selecting training set as balanced as possible and using the remaining as test set. Both RF and GBT made use of alpha and non-alpha samples, whereas the AE network discarded the negative (non-alpha) class.  

\begin{table*}[t]
\caption{F-score for different electrode set-ups, using alpha and non-alpha as positive class}
\label{table:results}
\centering
\begin{tabular}{|c||c|c|c|c|c|c|}
\hline
Positive class & CzOz & all & noCz & wearable & refT7 &Fp1Fp2\\
\hline
Alpha & \multicolumn{1}{r|}{0.94$\pm$0.06} & \multicolumn{1}{r|}{0.82$\pm$0.11} & \multicolumn{1}{r|}{0.81$\pm$0.12} & \multicolumn{1}{r|}{0.78$\pm$0.21} & \multicolumn{1}{r|}{0.74$\pm$0.26} &\multicolumn{1}{r|}{0.71$\pm$0.24}\\
Non-alpha & \multicolumn{1}{r|}{0.84$\pm$0.14} & \multicolumn{1}{r|}{0.85$\pm$0.13} & \multicolumn{1}{r|}{0.78$\pm$0.17} & \multicolumn{1}{r|}{0.71$\pm$0.27} & \multicolumn{1}{r|}{0.72$\pm$0.18} &\multicolumn{1}{r|}{0.72$\pm$0.16}\\
\hline
\end{tabular}
\end{table*}

\subsection{Results}\label{subsec:results}
Table~\ref{table:results} reports the F-scores obtained by the AE-based network using different EEG channel set-ups. As a reference, we report results with the CzOz channel. 
By comparing the F-scores, the NoCz configuration has a good performance (F-score=0.81), close to the all combination. Although the NoCz configuration demands 3 channels, corresponding to 5 electrodes, it is an affordable solution in view of a comfortable design, as it avoids the Cz electrode that is impractical (Fig.~\ref{fig:eeg}). The wearable configuration gives as well good results (F-score=0.78), despite a higher variance. Additionally, we report results using non-alpha as the positive class with no significant difference w.r.t. the model trained with alpha. This indicates that one-class training can be done using the class which is easiest to acquire in large quantities. 

\begin{figure}[t]
    \centering
    \centerline{\includegraphics[width=1\columnwidth]{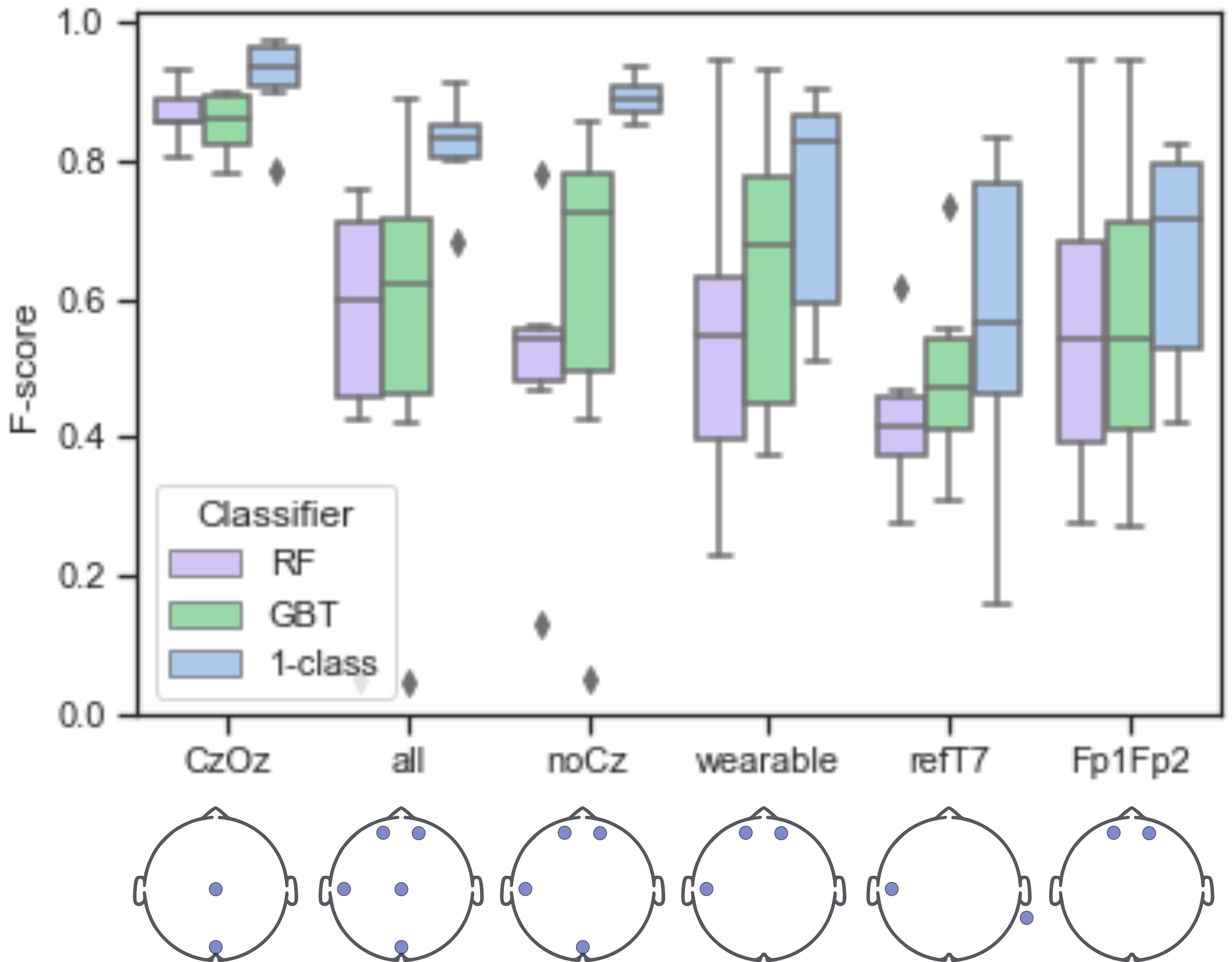}}
    \caption{F-score from the 6 experiments performed with Random Forest (RF), Gradient Boosted Trees (GBT) and 1-class methods, across all channels set-ups.}
    \label{fig:results}
\end{figure}

Figure~\ref{fig:results} compares the use of the AE-based architecture to RF and GBT. The AE-based network shows a consistent superior performance. These results are in-line with the demonstrated power of neural networks to learn latent representations and indicate that the AE-based network is a more reliable tool for EEG event detection through an optimized set-up.   

\section{DISCUSSION AND CONCLUSIONS}
We presented a one-class autoencoder-based framework to identify an optimal electrode set-up to detect and monitor events from EEG recordings.  We define the optimal set-up in terms of number of electrodes, comfortable location and event detection/monitoring performance. Using alpha wave detection as a use case, we investigated five electrode set-ups. The results indicate that the NoCz (F-score=0.81) and wearable  (F-score=0.78) set-ups are viable solutions for realistic monitoring of the studied use case. Although our results indicate that the proposed architecture is able to detect the presence/absence of alpha without using the CzOz channel, we observe that there is an increased variance in the performance as the number of channels is reduced. This suggests three things. First, that avoiding the most performing, but least comfortable channel (i.e. CzOz) comes at the cost of having to increase the number of alternative used channels. Second, that there is a minimal number of such alternative channels required to guarantee an acceptable performance. Third, that it is worth to explore additional set-ups, including other comfortable electrode locations, and/or other electrophysiological measurements.    
   
Our work is closely related to the more general problem of feature selection~\cite{alotaiby2015,balandong2018,chambon2018}, which accounts to select the EEG channels achieving the highest performing accuracy. In the scope of sleep studies, some methods have explored comfortable EEG channels, e.g. forehead electrodes, among the pool of features, reporting a performance accuracy of 76-77\%~\cite{chambon2018,huang2013b}. Our work differs from these previous approaches in two ways. Firstly, these methods focus on performance accuracy. As our final aim is to enable wearable EEG design, our optimal set-up definition goes beyond performance, comparing multiple set-ups considering  the comfort and discreteness criteria. 
Secondly, from a pure technical perspective, our one-class formulation avoids the problem of class imbalance, common to EEG analysis~\cite{chambon2018}. Although not directly comparable, this formulation allows our method to achieve a higher accuracy (83\%) than that one reported in previous works, in forehead electrodes.  

This work is a proof-of-concept on how learning-based techniques can assist the conception, development and implementation of every-day use wearable devices, which may go beyond the presented use case. Nevertheless, this framework represents just a small component of the infrastructure required to conceive a wearable device using an optimal set-up. For instance, alpha waves monitoring with the identified set-up needs to be done in conjunction with the AE network. Whether a light-weight AE network is to be deployed in the wearable itself or in a remote device (i.e. a phone) that does the processing is an important design choice that remains to be solved to be addressed as part of our future work.

\addtolength{\textheight}{-12cm}   

\end{document}